\documentclass[12pt]{JHEP}
\usepackage{amsmath,epsfig}
\usepackage{pb-diagram,lamsarrow,pb-lams,amsfonts,amssymb,amsthm}
 
\def\be{\begin{equation}}
\def\ee{\end{equation}}
\def\ba{\begin{eqnarray}}
\def\ea{\end{eqnarray}}

\def\be{\begin{equation}}
\def\ee{\end{equation}}
\def\bea{\begin{eqnarray}}
\def\eea{\end{eqnarray}}

\def\yzero{\smash{\hbox{$y\kern-4pt\raise1pt\hbox{${}^\circ$}$}}}
\def\p{\partial}

\def\be{\begin{equation}}
\def\ee{\end{equation}}
\def\bea{\begin{eqnarray}}
\def\eea{\end{eqnarray}}

\def\-{\hphantom{-}}

\def\s2{\frac{1}{\sqrt2}}

\def\beq{\begin{equation}}
\def\eeq{\end{equation}}
\def\beqa{\begin{eqnarray}}
\def\eeqa{\end{eqnarray}}

\def\IF{\relax{\rm I\kern-.18em F}}
\def\II{\relax{\rm I\kern-.18em I}}
\def\IP{\relax{\rm I\kern-.18em P}}
\def\IC{\relax\hbox{\kern.25em$\inbar\kern-.3em{\rm C}$}}
\def\IR{\relax{\rm I\kern-.18em R}}

\def\ch{{\cal H}}

\def\Dsl{\,\raise.15ex\hbox{/}\mkern-13.5mu D} 
\def\IZ{Z\kern-.4em  Z}

\title{A New Mechanism for Gauging a Theory}

\author{M. P. Garc\'\i a del Moral,\footnote{E-mail:
\emph{garciamormaria@uniovi.es}}\\
Departamento de F\'\i sica, Universidad de Oviedo, Avda Calvo
Sotelo S/n. Oviedo, Espa\~na}

\abstract{We provide a mechanism of gauging a theory based on a particular way to embed a theory on a target space such that a nontrivial fibration is produced. A connection over a nontrivial fibration with monodromy provides a natural framework for a new way of gauging a theory. Moreover, properties of the global symmetry of the original theory are included in a particular way in the new theory. This mechanism for gauging a symmetry preserves the total number of degrees of freedom in distinction with the classical one. We consider a particular example to illustrate the mechanism: by reinterpreting the supermembrane with central charges as a gauged supermembrane of the compactified supermembrane according to this new sense of gauging. Further applications are also  discussed.}

\preprint{FPAUO-11/08}
\begin{document}
\section{Introduction} Gauging is a powerful mechanism by means of which a global symmetry $G$, or a subset of it $H\in G$, is promoted to a local one in a new theory that typically preserves  a subset of the original global symmetry plus a new local symmetry. Electromagnetism and Yang-Mills theory are very well-known examples that can be obtained from a scalar theory and a Maxwell theory, invariant under global U(1) and SU(N) symmetries respectively. Generically new terms are required to be added by iterative processes to guarantee the invariance of the action found. In gauged theories typically matter is forced to interact with the explicitly introduced gauge field through the covariant derivative, in order the action to recover the symmetry invariance under the new gauge symmetry. This mechanism is also responsible for generically giving mass to scalar fields. Also on general grounds, the gauge symmetry, if present in the original lagrangian gets enhanced to a new one, when the former action couples to the new local symmetry that we have imposed. These way of finding new theories have provided a very useful arena in the context of maximal supergravities leading to gauged supergravities. Maximal supergravities in diverse dimensions $d<11$ appeared when 11D supergravity is compactified on a $T^d$, as found in \cite{cremmer-julia}. According to \cite{bl}, the mechanism of gauging was a systematic process introduced by Emmy Noether in which a lagrangian containing global symmetries gets deformed into a new one with those symmetries now elevated to local ones. This mechanism has being widely applied in physics. Supergravity theories, are gauged theories of lagrangians with rigid supersymmetries, that become deformed into new ones with local supersymmetries. Nowdays probably the more extended use of the gauging mechanism has being to obtain Gauged Supergravity theories in dimensions $d\le 9$, see a seminal paper by \cite{dwn} for compact spaces, also developed for noncompact spaces  in \cite{hull}. Gauged Supergravities share all the properties mentioned above and typically also lead to the presence of nonabelian gauge groups, give mass to the moduli stabilizing many of them, and allow for constructing theories with less number of supersymmetries. For a review and references therein, see \cite{samtleben}. Gauged supergravities can be obtained by compactifying maximal supergravities in upper dimensions $D$, on manifolds $d<D$ with nontrivial holonomy, by adding fluxes, twisting among other ways. So far, its interest lie in the fact that they have more chances to describe the physics in the low energy limit in the realm of elementary particles (string inspired supergravity models) and also in the context of string-inspired cosmological models. Gauged supergravities are thought to be connected with string/M-theory  theories in the presence of a nontrivial flux, although direct derivation -to the best of our knowledge- still has not been done. \newline

  In this note we are mainly interested in providing a different way to obtain a "gauged" theory understood in the following sense: The new theory contains a subset of the global symmetry included in the original theory, and a new gauge field is produced by the "gauging"  procedure, which consists in the following:  instead of adding a new gauge field by promoting the global symmetry to a local one, the gauge field it is going to be "extracted" from the closed forms present in the undeformed theory, by an appropriate handling of the global harmonic forms. Metaphorically is the analogous to what happens when we want to make an sculpture, one can model it by adding clay or extracting the figure from, let say, a marble with an appropriate chisel. To this second way of 'sculpting' lagrangians from an original one is the new gauging mechanism we propose. This mechanism is more restrictive -at least in its present formulation-  than the standard gauging mechanism originally introduced since it is strongly dependent on the physical properties of the initial action for obvious reasons: as happens in nature, 'extracting' is always more restrictive than 'adding'. In the context of String/M theory the interest of this new mechanism of gauging relies in the fact that it may be the key to solve the puzzle about the M-theory origin of gauged supergravities. Indeed, the Noether one applied in M-theory has not given the answer in these terms. For the M2 brane we propose the following diagram that relates with the two types of gaugings: 'Sculpting' and Noether and their relations:
\begin{equation}
\begin{diagram}
\node{\textrm{Compactified M2($n=0$)} } \arrow[2]{e,t}{\textit{'Sculpting'}}
     \arrow{s,l}{\textrm{Low Energies}}
        \node[2]{\textrm{M2 with central charges ($n\ne 0$)}} \arrow{s,r}{\textrm{Low Energies}}\\
\node{\textrm{Maximal Supergrav. (d)}}\arrow[2]{e,t}{\textit{Noether}}
   \node[2]{\textrm{Gauged Supergravities (d)}}
		\label{eq:diag2}
\end{diagram}
\end{equation}

  So the M2 brane with central charges discussed its construction in several papers \cite{mrt}-\cite{gmmpr} is an example of a gauged theory in this new sense, and we claim it corresponds to the M-theory origin of  nine dimensional gauged supergravities. This statement was already commented in \cite{gmmpr}, where the global description of the M2 brane with central charges in terms of symplectic torus fibrations with nontrivial monodromy  is shown and the precise details of these relation are going to appear in \cite{gmpr2}.

  This mechanism is discussed in section 2, in a more abstract formulation, and in section 3, it is the change of the symmetries involved.
  In section 4 we illustrate how this mechanism works. Then we will consider a precise example recently formulated in its global form \cite{gmmpr} in the context of M-theory: the supermembrane with central charges but now reinterpreting it as a gauged theory in this 'sculpting' sense. The supermembrane with central charges are interesting theories due to it quantum properties at supersymmetric level. They have a purely discrete supersymmetric spectrum at regularized level of the theory  can be interpreted as first quantized theories in distinction with the general case of supermembrane theories \cite{dwhn}-\cite{dwpp} that cannot be defined microscopically. Its spectral properties as well as its physical characteristics have been extensively studied in \cite{gmr}-\cite{bgmr3}. We will compare them with the ungauged theory of the supermembrane compactified on a 2-torus, as the simplest formulation.  We will see how it shares some of the typical properties of a gauged theory while not all: only part of the former global symmetry survives, a new gauge symmetry is introduced but instead of enhancing the former gauge theory, reduces it: the infinite group of symplectomorphisms preserving the area of the 2-torus are restricted to those preserving the structure group of the fiber with a particular monodromy contained on the $SL(2,Z)$ mapping class group of the torus, for example a $Z_2\times Z_2$, supersymmetry is spontaneously broken from $N=2$ to $N=1$, and scalar fields acquire a mass \cite{joselen}.
We discuss in a qualitative way in section 5, on possible applications of this mechanism to other theories, for example p-brane theories formulated in terms of Poisson-Nambu brackets associated to compactified sectors restricted to have some monopole contributions. In section 6 we conclude and also comment on the relation between the supermembrane  with central charges as a gauged theory and its relation with gauged supergravities in the low energy limit. The precise realization and details of this statement are going to apppear in \cite{gmpr2}. In this paper we just illustrate the way this mechanism works.

\section{The mechanism}
In order to be selfcontained, let us briefly remind the simplest example of what we usually understand by a gauging mechanism:
\subsection{Noether gauging mechanism for an abelian symmetry}
According to \cite{bl} book, Noether mechanism is a systematic technique for deriving an action with a  local symmetry from an action with a global symmetry. To illustrate it we will sketch it closely following standard references. The simplest example of this procedure is the action of a massless Dirac field $\Psi$,
\bea S_0=i\int d^4 x \overline{\psi}\gamma^{\mu}\partial_{\mu}\psi.
\eea This action is invariant under the transformation \bea \psi\to
e^{-i\epsilon}\psi \eea where $\epsilon$ is a constant phase, so the
symmetry of $S_0$ is global. To make it local the parameter has to
depend of the point of the space-time, that is  $\epsilon(x^{\mu})$.
However the former action is not invariant under the local
transformation \bea \psi\to e^{-i\epsilon(x)}\psi \eea and changes by
an amount \bea \delta S_0=\int d^{4}x
\overline{\psi}\gamma^{\mu}\psi\partial_\mu \epsilon= \int d^{4}x
j^{\mu}\partial_{\mu}\epsilon \eea where
$j^{\mu}=\overline{\psi}\gamma^{\mu}\psi$. To restore the invariance
a gauge field $A^{\mu}$ is introduced that transforms as \bea
A_{\mu}\to A_{\mu}+\partial_{\mu}\epsilon(x^{\mu}) \eea So a term coupling
the Noether action is added to make it invariant \begin{equation}\begin{aligned}S_1&=S_0-\int
d^4xj^{\mu}A_{\mu}=\int d^4
xi\overline{\psi}\gamma^{\mu}(\partial_{\mu}+iA_{\mu}\psi)\\
&= \int d^4
xi\overline{\psi}\gamma^{\mu}D_{\mu}\psi\end{aligned}\end{equation}
There are also terms for the kinetic term of $A_{\mu}$ and the
electromagnetic coupling constant is absorbed in the definition of
$A_{\mu}$.
 The final gauged theory is invariant under the former global $U(1)$ symmetry and the new gauge $U(1)$ symmetry. For  the case of a initial nonabelian global symmetry case, a locally invariant final action requires of a finite step iterative process of adding further terms to the transformation law and to the action.
\subsection{Non-trivial fibration mechanism}

Let us review the basics of a nontrivial fiber bundle properties firstly. Consider a non trivially fibered manifold $E$ consistent on a base manifold $B$, a fiber $F$, a group $G\subset Aut(F)$, an open cover $\{U_i\}$\footnote{For physical theories this open cover can be assumed to be  a good cover $\{\underline{U_{\alpha}}\}$ in the Cech cohomology sense.} of $E$ and transition functions $t_{ij}$ (one-cocycle in $Z^1 (B,Aut(F))$) defined on them. We will assume that the base manifold $B$ is compact\footnote{For the noncompact base manifolds one has to guarantee the non triviality of the fiber bundle $E$.}. When the manifold is nontrivially fibered we cannot write it as the simple product of $E=B\times F$. This means that we need several charts $\{U_i\}$ to cover it  such that the transition functions exists. The different fibrations for a given base $B$ and a given fiber $F$ are classified by the corresponding characteristic classes. (Take for example the case of line bundles classified by the Chern classes  and labeled by integers $n\ne 0$, being $n=0$ corresponding to the trivial fibration).  So far, we only have imposed to our manifold to be a fiber bundle; that is, it possesses local trivializations $\phi_{i}: E_{U_i}\to F\times U_i$ associated to the different charts $U_i$ and transition functions $t_{ij}:\phi_j\circ \phi_i^{-1}: U_i\cap U_j\to Aut(F)$ relating the different patches. Take for example the magnetic monopole case with the transition function mapping the equator $S^1$ to the $U(1)$ fiber through an integer $n$ that classifies those fibrations\footnote{Rigourously fibrations are generalizations of the concept of fiber bundle and do not admit always local trivializations. They are continuous maps between topological spaces $\pi:X\to Y$ such that they satisfy the lifting homotopic property from $X \to Y$. Here, and along the text we will use these concepts indistinctly unless otherwise signalled.}.

We can introduce a connection on the fibre bundle, this is an extra object that we may add or not to it. Connections are defined globally on $E$ although they also carry local information. Indeed, connections allow comparison between different fibers associated with
different (distant) points of the manifold. Physically they are more commonly known as gauge potentials. Let us define $s_i:U_i\to E$ the local sections ($\pi\circ s=id_U$) and $\omega$ is the Ehresmann connection defined globally over the bundle $P(B,G)$. Let be $A_i$ the connection on a trivialization satisfying the following condition
 \bea\label{u}
A_j=t_{ij}^{-1}A_it_{ij}+t_{ij}^{-1}dt_{ij}.
\eea
Since a non-trivial principal bundle does not admit a global section, the pull back $A_i=s_i^* \omega$ exists locally but not necessarily globally.  The one-form connection on the principal bundle $P(B,G)$ allows to define 'horizontal subspaces'. The connection one-forms $A_i$ are associated to the trivial bundle $\pi^{-1}(U_i)$ and do not have global information on $E$, so it is needed the complete set $\{A_i\}$ to define it globally on $E$. That is the set $\{A_i\}$ have the information through the transition function about the global symmetries of the theory, this means that if we define a covariant derivative on the associated bundle, (for example a connection on  a Weyl bundle), it contains local information of the manifold but in addition, it also contains information about the global symmetries that allow to patch the different charts of the manifold. So summarizing so far, the introduction of a connection on the nontrivial fiber bundle allows to realize local symmetries and to define covariant derivative on associated vector bundles which  also have information about the global symmetries of the theory via the nontrivial patching. The action $\mathcal{S}$ of the theory is the functional gauge invariant of the sections. Given a particular fibration it can be constructed several actions and respectively lagrangians $L$, compatible with the same sections for a given fiber bundle.\newline
\vspace{5mm}

\emph{\bf General deformation of a Fiber bundle}.  For classifying fibrations it is particularly useful the Cech Cohomology  group over integers, $H^*_C(B,\mathbb{Z})$, since in distinction with DeRham Cohomology of closed forms $H^*_{DR}(B,\mathbb{R})$, with which is isomorphic on the real-valued cocycles, it is sensitive to the classification of torsion elements present on a fibration. The deformation of the fibre bundle we will consider in principle is quite general. Indeed, the word 'deformation' used here, does not necessarily implies infinitesimal deformation a priori, but a change in the cohomological class of the fibration due in some cases to homotopical changes of the base\footnote{See \cite{alvarez} for a discussion about the relation between cohomology and topology.}, base and fiber, or of the complete fibration. Consider the undeformed fibration, $F_0\stackrel{i_0}{\rightarrow} M_0\stackrel{\pi_0}{\rightarrow} B_0$ which  can be trivial or not, (in principle we assume it completely general), without any requirement of having a 1-form connection $A_i$ on it, and a completely general deformed fiber bundle, $F_1\stackrel{i_1}{\rightarrow} M_1\stackrel{\pi_1}{\rightarrow} B_1$. We assume changes in the topology of the fibration such that cohomology is changed, which in most of the cases it also implies changes in the homotopy class.
We define the following deformation of parameter spaces associated to two inequivalent cohomology classes: $\{B_0,F_0,G_0,\pi_0,t_{ij},s_i\}$ in $ H^*_0(B_0,\mathbb{Z})\to \{B_1,F_1,G_1,\widetilde{\pi}_0,\widetilde{t_{ij}},\widetilde{s_i}\}$ in $H^*_1(B_1,\mathbb{Z})$. The relation between the transition functions and sections old and new is the following one\footnote{We thank A. Vi\~ na for explaining this to us.}
 \bea
 \widetilde{t}_{ij}=t_{ij}+\xi_{ij}\quad \textrm{and}\quad
 \widetilde{s_i}=s_i+\Sigma_i
 \eea
   satisfying both fibrations the old one and the new (deformed) one, the standard cocycle conditions on the overlap of three coordinate patches and the consistency conditions for the sections,
 \bea\label{aa}
 t_{ij}t_{jk}t_{ki}=\mathbb{I} \quad \textrm{for}\quad U_{i}\cap U_j\cap U_k\ne\oslash,\quad s_i=\sum_j t_{ij}s_j.\\
 \widetilde{t}_{ij}\tilde{t}_{jk}\tilde{t}_{ki}=\mathbb{I} \quad \textrm{for}\quad \widetilde{U}_{i}\cap \widetilde{U}_j\cap \widetilde{U}_k\ne \oslash; \quad\widetilde{s_i}=\sum_j \widetilde{t}_{ij}\widetilde{s}_j.\\
 \eea

 Then, one can express the deformation in one section in terms of the remaining new sections, their deformation and that of the transition functions.

 \bea
\Sigma_i=\sum_j (t_{ij}\Sigma_j+\xi_{ij}\widetilde{s}_j)=\sum_j (\widetilde{t_{ij}}\Sigma_j+\xi_{ij}s_j)
 \eea
together with the cocycle consistency condition
\bea
t_{(ij}t_{jk}\xi_{ki)}+t_{(ij}\xi_{jk}\xi_{ki)}+\xi_{(ij}\xi_{jk}\xi_{ki)}=0
\eea

For infinitesimal deformations those conditions reduces to
\bea
\Sigma_i=\sum_j (t_{ij}\Sigma_j+\xi_{ij}s_j);\quad t_{(ij}t_{jk}\xi_{ki)}=0.
\eea
 Consider $h_{i,p}$ is a map that change from one fiber to an \underline{inequivalent} one.
 \bea
 h: (E_0,G_0)\to (E_1,G_1)\\
 h_{i,p}=\phi_{i,p}^{-1}\widetilde{\phi}_{i,p}
 \eea
 with $\phi_{i,p}$ a local trivialization of the undeformed fibration $F_0\to M_0\stackrel{\pi_0}{\rightarrow} B$
and $\widetilde{\phi_{i,p}}$ a trivialization of the deformed fiber bundle, $F_1\to M_1\stackrel{\pi_1}{\rightarrow} B$, satisfying the previous (\ref{aa}) consistency conditions. The deformation of the transition functions $\xi_{ij}$ can also be thought in the following equivalent way
 \bea
 \widetilde{\xi_{ij}}=h_{i,p}^{-1}t_{ij}h_{j,p}-t_{ij}
 \eea
Then
\bea
\widetilde{s_i}= h_{i,p}^{-1}t_{ij}h_{j,p}s_j+ h_{i,p}^{-1}t_{ij}h_{j,p}\Sigma_j
\eea

 The new sections corresponds to a deformation of the former ones together with a deformation on the fibration via the changing of the transition functions. A deformation of this type, in principle, may also imply a change in the degrees of freedom of the physical theory defined on it. Given the space of sections and determined the interactions among them, two actions can be defined associated to the undeformed and deformed fibration. We will denote $\mathcal{S}_0$ and $\mathcal{S}_1$ respectively.\newline

{\bf Interesting deformations}
The deformations so far discussed are the framework where this new gauging mechanism acts. To obtain relevant physics it can be more useful to restrict set of parameter space of deformations.We will concentrate on a particularly interesting type of deformations -without excluding further generalizations-, that provides a new mechanism of gauging. \newline
Before entering to describe it, we will briefly comment on a set of interesting topological deformations that have been widely discussed in the literature from the perspective of this scheme: we refer to the physical theories containing topological defects. One can understand the local approach of those theories (without including topological defects) in terms of trivial fibrations. The deformation in those theories  keep fix the sections $s_i=\widetilde{s_i}$, the structure group $G_1=G_0$, and the fiber $F_0=F_1$ but changes the topology of the base, so its homotopy class, meaning changing the compactness property and contractible properties of the base manifold $B_0\to B_1$. The flow to the new base $B_1$, allows for the existence of nontrivial transition functions. The deformed fibration that is chosen in these theories corresponds to have a it nontrivially fibered, constructed in terms of $\widetilde{t_{ij}}$, and projection $\widetilde{\pi_i}$ but imposing that the structure group of the fiber  $G$ is preserved. The space of deformation parameter is $\{B,t_{ij},\pi\}$. This deformation is not soft, in the sense that deformation is not infinitesimal. It corresponds to the well-known formulation of a physical theory as a non trivial fibration departing from the trivial fibration. Consider for example the deformation of a Yang-Mills theory over $R_2\times SU(N)$ with $t_{ij}=\mathbb{I}$ to a Yang-Mills theory with monopoles, for example by taking the base $S^2$  and a keeping constant the structure group $SU(N)$ of the fiber but imposing it to be non trivially fibered, then, a set of $\{s_i\}$ is needed to define the 1-form connection. It corresponds to have
\bea
s_i= \widetilde{t_{ij}}s_j\quad \textrm{with} \quad \Sigma_j=0,
\eea
together with the cocycle condition.
 This illustrate schematically from this global point of view, the change in fibrations of a useful procedure that has been extensively used to  introduce global solutions as monopoles, vortices, instantons, among others, very relevant for quantum theories. This type of approach has been exhaustively studied in the literature and we will not concentrate here further on it since it does not provide by itself a mechanism of gauging.  \newline
\subsection{The New Gauging Mechanism} The new gauging mechanism consists in the deformation of fibrations keeping the homotopy-type of the base $B$ and of the fiber $F$ unchanged, but allowing changes in the complete fibration $E$\footnote{The base and the fiber changes in a number of ways for example in the associated moduli space due to the nontriviality condition, however homotopy-type in both spaces is preserved.}. The sections and the structure group change together with the deformation parameters. The space of deformation parameter $\{s_i,t_{ij},G,\pi\}$.
A natural set-up for these type of gaugings consists in theories whose base manifold is a compact p-dimensional manifolds and whose fiber space corresponds to the target space compactification manifolds, although more general realizations could also be considered. The trivial fibration in these cases corresponds to ordinary compactifications, and the deformed one corresponds to twisting the fiber space associated to nontrivial compactifications in a particular way.

Take for example, the particular case in which the undeformed fibration is trivial $t_{ij}=\mathbb{I}$ and the 1-form connection defined on it is also trivial. Then,
\bea
\widetilde{t_{ij}}\simeq \xi_{ij}= g_{i,p}^{-1}g_{j,p}
\eea
with $g_{i,p}^{-1}g_{j,p}=h_{i,p}^{-1}h_{j,p}-\mathbb{I}$.
Assume for simplicity the following: only one type of sections gets deformed  and it corresponds to the gauge fields. Assume also that the original theory only contains scalar fields that remain invariant, so for gauge fields that we will concentrate on $s_i=0$. By modifying the cohomology class $H^2(B,\mathbb{Z})$ through some appropriate topological condition, (we give some examples of topological conditions along the next sections). For the set-ups considered here, they are associated to conditions on the embedding of the worlvolume manifolds on the compactification space, but probably could also be generalized. It produces a deformation $\Sigma_j$ and new sections appears $\widetilde{s_i}$ then,
\bea
\widetilde{s_i}\simeq \Sigma_j;\quad \xi_{ij}\xi_{jk}\xi_{ki}\simeq \mathbb{I}.
\eea
The structure group of the deformed fibration $G_1$ becomes changed with respect the one of the former structure group $G_0$ since the new transition functions $\widetilde{t_{ij}}$ get restricted by the consistency condition. In a trivial fiber bundle all 1-forms are connections and globally defined, in a nontrivial fibration only a subset of those satisfy (\ref{u}). With those one-forms is possible to define globally a connection such that its curvature is given by a topological condition. To realize them as deformation of theories with closed one-forms, it is needed an appropriate handling of the harmonic contributions. A more clear explanation is given along the next sections where a concrete realization is provided.
There are obvious generalizations of this procedure by allowing deformations of all the different types of sections (scalar fields, spinorial fields, etc.. with $s_i\ne 0$ for each of them and considering as undeformed fibration any kind of them, not restricted to the trivial one ($t_{ij}\ne \mathbb{I}$).
\section{Symmetry Change in the New Gauging Procedure.}
In this section we want to study this deformation from the physical point of view, by characterizing the effect of the twisting in the symmetries of the gauged Lagrangian $L_1$ associated to the deformed fibration $E_1$.
\subsection{Changes in the global symmetries} Let us consider a subset of the global symmetries that a given undeformed theory can contain: Let us concentrate on those naturally included in this set-up via the nontrivial patching of the different charts which allow to define globally the connection. We will consider Lagrangians that realize these symmetries. The cocycle conditions associated to nontrivial patching represents extra restrictions with respect to the undeformed case. Because of it, the global symmetry group $H$ associated to the base of the deformed fibration is a subgroup of the global symmetry group $G$ of the undeformed fibration\footnote{For the case of arbitrary deformations this is not necessarily true, since it can be that not even a subgroup of the former global symmetry is preserved and it may happen that new global symmetries may emerge at the level of the deformed Lagrangians.}. Since these global symmetries are associated to the nontriviality of the transition functions due to the patching of the base manifold, there are isometries naturally associated to the image of the deformed base manifold in the fiber (target space) $I$. There are also relevant discrete symmetries that appear as subgroups $\Gamma\subseteq (H(\mathbb{Z})$ with $H(\mathbb{Z})$ the Large Diffeomorphism group of the compact base manifold $B$, the Mapping Class Group MCG\footnote{The $MCG=Diff(B)/Diff_0(B)$ with $Diff_0(B)$ the infinite group of Diffeomorphisms connected to the identity. For example $MCG(T^n)\cong GL(n,\mathbb{Z})$. The Farrell cohomology of $GL(n,\mathbb{Z})$ was computed in \cite{ash}, generalized to more general mapping class groups in \cite{gmx}.}. The homotopy group of $Aut(E)$ is the Mapping Class Groups (MCG(B)) of the base manifold. In the case of marked Riemann surfaces $\Sigma_{g,r}$ they are generated by Dehn twists. The mapping class group is generated by elements of torsion\footnote{The $MCG(\Sigma_{g,r})$ with $\Sigma_{g,r}$ a Riemann surface of genus $g$ with $r$ marked points, is generated by torsion elements whenever $(g,r)\ne(2,5k+4)$ with $k$ an arbitrary integer \cite{feng-luo} }. The deformation of the fibration, by twisting the base and fiber topological spaces chooses a class representative of the MCG(B). It induces a monodromy on the fibers due to the nontrivial class of torsion elements in the cohomology class \cite{khan}, see also \cite{walzack}. The monodromy representative appears naturally in the context of multivalued functions. It is defined by measuring what happens in the fiber as we give a loop around a point $b\in B$ of the fundamental group $\pi_1(B,b)\to S(F)$ with $S$ the symmetric group of permutations\footnote{More formally $\rho: \pi_1(N(\mathbb{U}))\to Aut(G)$ is the map of the fundamental group from the Nerve of the good cover $\underline{U}$ on a connected topological space $B$ to the automorphism of the structure group of the fiber $G$.}.\newline


\subsection{Changes in the gauge symmetry} It is relevant for the gauging mechanism to see how the deformation of the fibration induces new sections, gauge fields, in the physical theory associated to the deformed fibration. Assume a class of examples consisting in theories containing closed one-forms, over a compact base manifold. We analyze in detail one of these examples in the next section, here we give the guidelines for a general setting. The pullback of the connection on the compact base manifold is a one form that can be decomposed by using the Hodge theorem in its exact, co-exact and harmonic pieces,
 \bea
 dX= d\alpha+d*\beta+h
 \eea
 with $d\alpha=0$ the exact part and $d*\beta=0$ the co-exact piece and $h$ are the harmonic forms.
 Whenever we deal with a compact base manifold with 1-cycles or singularities for which, some one-form closed operator can be defined  and the operator has a nontrivial harmonic piece. The exact piece emerging from the decomposition is the one-form connection associated to the principal fiber bundle for those manifolds where the following topological restriction can be imposed,
 \bea \label{a}
 \int_{\Sigma_2}F_2=n\in \mathbb{Z}\quad n\ne 0.
 \eea
That is, there is a  $\Sigma_2$ is a 2-cycle contained in the compact base manifold $B$.
\newline

 By Weil's Theorem
\cite{weyl}, there exists a U(1) principle bundle and a connection over it such
that its pullback by sections over $B_1$ are 1-form connections with curvatures $F_r=dA_r$ labeling $r$ the number of gauge fields,
given by (\ref{a}). The gauge symmetry of the physical theory of the associated vector bundle $\mathcal{A}_r$ can be however more complicated, since the undeformed theory $L_0$ may originally posses a gauge symmetry, take for example the $Diff_0^+B_0$. The monodromy contribution imposes restrictions in the type of Symplectomorphims $Sym(B_1)$ compatible with the symplectic structure $\omega_F$ of the twisted fiber by the monodromy representative, such that the effective gauge symmetry of the 1-form connection $\mathcal{A}_{r}$ on the associated vector bundle is $G_1=Diff_0^+(B)\vert_{_{MCG}}$. In general $G_1\subset G$, with the restriction associated to the $\Gamma\subseteq H$ global symmetry of the harmonic forms.
Finally fixing the global symmetry of the theory by means of the deformation of the fibration,  fixes the harmonic sector and allows to introduce it in the definition of a global covariant derivative $D_r$\footnote{In order to understand properly this rotated derivative as a covariant derivative, one has to think of it as the derivative $D_{skew\nabla_r}$.} (rotated by the harmonic contribution as it is done in \cite{mr}), the new theory has a new degrees of freedom associated to the gauge field connection $\mathcal{A}_r$ preserving the structure group of the fiber and a covariant derivative $\mathcal{D}_r\bullet = D_r\bullet+ \{\mathcal{A}_r,\bullet\}$ where the monodromy and the harmonic contribution have already been incorporated in the definition of $D_r$. There is  a change in the types of degrees of freedom of the theory, where the $r$ closed 1-form are converted into the new $r$ gauge vector potentials but preserving the number of degrees of freedom (d.o.f.). See the example described in the next section for understanding better the details.

\subsection{Changes in the base and fiber manifold}
 The base manifold always possesses a natural group of gauge symmetry: the infinite group of diffeomorphism preserving the p-volume of the compact base manifold $B_p$ with $p$ denoting its spatial dimensions. However, this symmetry is not always realized at the level of the action $\mathcal{S}_1$. For those physical theories whose lagrangians are invariant under it, take for example p-branes, the 1-form connection of the principal bundle is inherited by the associated vector bundle with a group which corresponds to the restriction of the $Diff^+(B_0)$, the infinite group of diffeomorphims preserving orientation, by the global symmetries that fixes the harmonic sector.
 This deformed fibration  although homotopically invariant, has a new base $B_1$ with a new metric $g_1$  compatible with the symplectic form of the fiber and new isometry group $I_1$. The new gauge symmetry $\mathcal{A}_r$ is restricted to be a particular class of $Symp(B_1)$ labeled by the characteristic classes compatible with the new fibration $E_1$. The projection of the symplectic form on the base manifold $\omega_1$ is \emph{inequivalent} to the one of $B_0$, so it is the metric $g_1$ with respect to $g_0$. This implies a restriction of the gauge symmetry $Symp(B_1)\subset Diff^+ (B_0)$. This deformation also has effects on the original isometry group that becomes changed $I_0\to I_1$.  The deformation can be explicitly studied by comparing their respective killing vectors \cite{tesisjoselen}.
 Consider as base manifolds for example, Riemann surfaces. They are Kahler manifolds endowed with a complex structure form $J$ compatible with the symplectic structure $\omega_1$ defined on them, this means that the metric $g_1$ is defined in terms of the symplectic form and inherits the gauging properties of the new fiber defined by the new connection $\mathcal{A}_r$.

 We have said that this mechanism leaves invariant the topology type of the fiber (and the base), but the existence of the principal fiber bundle imposes a change also in the fiber, in the sense that the compatibility condition for the symplectic connection to exists can also modify for example the moduli space of the fiber or the isometry group $I_1$ via the inverse of the pull-back action in such a way that the complete fibration $E_1$ will admit a symplectic form. A more detailed analysis will be considered elsewhere.

\subsection{Changes in the Lagrangian } We consider two theories in which the topology type of the base $ B$ and the fiber $F$ are unaltered, but it changes the one associated to the fibration of the manifold $E$. We consider for a given set of sections $s_i$, global and local symmetries with groups  $H$, $G(x)$, for $x\in B$ respectively, and a given set of interactions, $i_0$,  $L_0(s_j; G_0(x),H;B,F;i_0)$. The twisting in the fibration produces a deformation \-
$L_1(s_{j-r},A_r;G_1(x),\Gamma;B,F;i_1)$ with $\Gamma\subset H$, and $G_1(x)=G_0\vert_{\Gamma}$, $i$ the total amount of d.o.f and $r$ the number of gauge connections induced in the associated fiber bundle. Indeed the topological condition modifies the lagrangian,
\bea
L=L_0+\textrm{Top}
\eea
The topological term produces the quantization condition for the existence of a principal fiber bundle. It also guarantees the cancelation of total derivative terms in the action producing changes in the allowed interactions $i_1$. For the case of the supermembrane explianed in next section, this difference implies the possibility of quantizing the hamiltonian of the theory in distinction with the $D=11$ case and in the standard compactified one. By rewriting the derivatives as a covariant derivative through a proper handling of the global symmetries, the new gauged Lagrangian $L_1(i_1)$  appears.

 %
\section{A example: The supermembrane with central charges.}

In \cite{gmmpr} the authors give the explicit formulation of the 11D supermembrane compactified on a torus with a topological condition (an irreducible wrapping condition)  called central charge condition \footnote{The name obeys to the fact that this condition guarantees the existence of a nontrivial central charge in the supersymmetric algebra \cite{mrt}\cite{dwpp}} as a symplectic torus bundle with non trivial monodromy and non vanishing Euler class. This
construction allows a classification of all compactified supermembranes on a torus showing explicitly the discrete SL(2,Z) symmetries associated to dualities. It hints as the origin in M-theory
of the gauging of the effective theories associated to string theories.

In particular in \cite{gmmpr} it is shown that the Supermembrane with central charges may be
formulated in terms of sections of symplectic torus bundles with a
representation $\rho:\pi_1(\Sigma)\to SL(2,Z)$ inducing a
$Z[\pi_1(\Sigma)]$-module in terms of the $H_1(T^2)$ homology group
of the fiber. The hamiltonian together with the constrains are
invariant under the action of $ SL(2,Z)$ on the homology group $H_1(T^2)$ of
the fibre 2-torus $T^2$. Geometrically to guarantee the existence of a symplectic form in the complete fiber space $E$ structure is only the possible
 if and only
if the characteristic class is a torsion class in
$H^2(\Sigma,Z_{\rho}^2)$ \cite{khan}, see also \cite{walzack}. Locally the target is a product of
$M_9\times T^2$ but globally we cannot split the target from the
base $\Sigma$ since $T^2$ is the fiber of the non trivial symplectic torus
bundle $T^2\to\Sigma$.

In this section, we want to understand this construction as a gauged theory under the light of this 'sculpting' mechanism. Let us take as the undeformed fibration the trivially compactified supermembrane  with worldvolume for symplicity $\Sigma_{1}\times R$, with $\Sigma_1$ representing a Riemann surface of genus 1, compactified on a target space $M_9\times T^2$.
The corresponding lagrangian in the L.C.G. is,
\bea\label{4} \mathcal{L}=P_{m}\dot{X}^{m}-\mathcal{H} \eea where
the physical hamiltonian is given by \bea\label{5}
H=&T^{-2/3}\Large\int_{\Sigma}\sqrt{W}[\frac{1}{2}(\frac{P_{m}}{\sqrt{W}})^{2}+
\frac{1}{2}(\frac{P_{r}}{\sqrt{W}})^{2}
+\frac{T^{2}}{2}\{X^{r},X^{m}\}^{2}\\ \nonumber &
+\frac{T^{2}}{4}\{X^{r},X^{s}\}^{2}+\frac{T^{2}}{4}\{X^{m},X^{n}\}^{2}]+\textrm{fermionic
terms} \eea subject to the constraints \bea \label{6}
 d(P_{r} dX^{r}+P_{m}dX^{m}+ \textrm{fermions})=0\eea\bea \label{6a}
& \oint_{\mathcal{C}_{s}}(P_{r} dX^{r}+P_{m}dX^{m}+ \textrm{fermions})=0. \eea $\mathcal{C}_{s}$ is a canonical
basis of homology on $\Sigma$. The constraints (\ref{6}),(\ref{6a}) are the
generators of area preserving diffeomorphisms homotopic to the
identity. The bracket in (\ref{5}) is given by \bea
\{X^{m},X^{n}\}=\frac{\epsilon^{ab}}{\sqrt{W}}\partial_{a}X^{m}\partial_{b}X^{n},
\eea it is the symplectic bracket constructed from the
non-degenerate two-form \bea \sqrt{W}\epsilon_{ab}d\sigma^{a}\wedge
d\sigma^{b} \eea over $\Sigma$. In 2-dim the area preserving
diffeomorphisms are the same as the symplectomorphisms.
Let us perform a deformation in the fibration as indicated in section 2, by keeping invariant the same topological base and fiber (the compactification manifold), but imposing a obstruction to the triviality called 'the central charge condition'. This is a condition that represents a twist in the fibration produces a principal fiber bundle. The lagrangian of the undeformed fiber has the following symmetries: a gauge symmetry  $DPA_0(t^2)$, target space susy $N=2$, a global symmetry $H\equiv Sp(2,Z)$ associated to the wrapping condition of the embedded maps $t^2\to T^2$:
 \bea
 \begin{aligned}
&\oint_{C_s} dX^r=n_s^r\in Sp(2,Z)\quad C_s \quad \textrm{the homological basis of}\quad T^2\\
 &\oint_{M_9}dX^m=0
 \end{aligned}
 \eea
{\bf Step 1}: Perform a Hodge decomposition of the closed forms in terms of harmonic one-forms $d\widehat{X}_r$ and a exact one-form $dA_r$:
 \bea
 dX_r=P_r^s d\widehat{X}_s+dA_r
 \eea
 {\bf Step 2}: Define a associated principal fiber bundle. For the case of the supermembrane it was found in \cite{mrt} to be related with imposing on the maps the following topological condition associated to an irreducible wrapping
 \bea
 \int_{\Sigma} dX_r\wedge dX_s=n\epsilon^{rs}Area_{\Sigma}
 \eea
 This condition is also associated to the presence of a nontrivial central charge in the supersymmetric algebra and because of this, the associated theories have been named: {\tt The supermembrane with central charge theories}.
 This condition corresponds to the presence of monopoles on a Riemann surface $\Sigma_g$ with genus larger or equal to one $g\ge 1$ found in \cite{monopole}. For the $g=0$ case it reduces to the well-known Dirac monopoles, however that case does not interest us for our present purpose since there are no harmonic one-forms in $S^2$, so  we will restrict to the MR-monopoles\footnote{We are calling MR monopoles to the generalization of Dirac monopoles to Riemann surface with arbitrary genus discovered by I. Martin, A. Restuccia. These canonical monopole connections are solutions of the projected Seiberg-Witten equations over compact Riemann surfaces.}.
  The harmonic one-forms due to the wrapping condition have a global Sp(2,Z) symmetry of the mapping class group.
  As a consequence of the nontrivial fibration the values of the matrix are: \bea P_r^s= M_{r}^s= 2\pi R^r S_r^s\quad with \quad S_r^s\in Sp(2,Z)\eea
  {\bf Step 3}: Associate a one-form connection to the nontrivial fiber bundle. We define a symplectic connection $A$ preserving the structure of the fiber under holonomies. To this end, first we define a rotated derivative associated to the Weyl bundle \cite{mr}:
  \bea
  D_r\bullet=( 2\pi R^rl^r)\frac{\epsilon^{ab}}{\sqrt{W(\sigma)}}\partial_a\widehat{X}^r(\sigma)\partial_b \bullet
  \eea
  At the moment we define this rotated derivative, we are performing an \textit{extension of the covariant derivative definition}, in which the associated bundle has a nontrivial monodromy.
  The related derivative fixes a scale in the theory and breaks the former $H= Sp(2,Z)$ theory to a subgroup $\Gamma\in Sp(2,Z)$ by specifying the integers of $S_r^s$. Fixing $R^r$ also fixes the Kahler and complex structure geometrical moduli when expressed in complex coordinates.
  The symplectic covariant derivative is then:
  \bea
  \mathcal{D}_r\bullet=D_r\bullet+\{A_r,\bullet\}
  \eea
  and then the connections  $A_r$ transform with the symplectomorphim like:
  \bea
  \delta_{\epsilon}A_r=\mathcal{D}_r\epsilon
  \eea
  I will call from now on, a $\Gamma$-invariant connection of symplectomorphism.\newline
 {\bf Step 4}- Project this one-form connection onto the base $\pi: A\to \Sigma_g$. The associated symplectic form is
  \bea
  \omega_{ab}=Sech(\sigma^b)^2(d\sigma^a\wedge d\sigma^b)
  \eea
  This symplectic form is clearly inequivalent to the canonical one associated to the flat torus $t^2$ considered for the undeformed fibration (the trivial one). This means that the nontrivial fibration plus the handling of global forms implies a deformation in the base manifold, indeed the isometry group closely related to the harmonic group of symmetry is not the associated to a flat torus.
 Since a Riemann manifold has three compatible structures $g_{ab},J,\omega_{ab}$ the metric is associated to the harmonic one-forms that preserve the fiber associated to the MR-monopoles, the induced symplectomorphism do not lie in the same conformal class of the flat torus. There is a
compatible election for $W$ on the geometrical picture we have defined.
We consider the $2g$ dimensional space of harmonic one-forms on $\Sigma$. We
denote $dX^{r}$, $r=1,2$, the normalized harmonic one-forms with
respect to $\mathcal{C}_{s}$, $s=1,2$, a canonical basis of homology
on $\Sigma$: \bea \label {7}
\oint_{\mathcal{C}_{s}}d\widehat{X}^{r}=\delta^{r}_{s}. \eea We
define \bea \label{8}
\sqrt{W}=\frac{1}{2}\epsilon_{rs}\partial_{a}\widehat{X}^{r}\partial_{b}\widehat{X}^{s}\epsilon^{ab},
\eea it is a regular density globally defined over $\Sigma$. It is
invariant under a change of the canonical basis of homology. This choice of the metric of the base manifold guarantee the compatibility between the symplectic forms of the base and the fiber.
\newline
In \cite{gmmpr} it is shown the formulation of the supermembrane in terms of
sections of the symplectic torus bundle with a monodromy  is a nice geometrical structure to analyze global aspects of gauging procedures on effective theories arising from M-theory. We noticed the particular case in which the representation $\rho$ is given by the matrix
 \bea\label{182}
\begin{pmatrix} 0 & 1\\
                        -1& 0
 \end{pmatrix}^{M+N}
 \eea
 the subgroup reduces to $Z_2\times Z_2$ and this case was considered in several papers \cite{mor}-\cite{g2}. A detail analysis of this particular example is going t appear in\cite{gmmpr2}.
 \newline
There are formal details in this particular construction are going to appear in an independent paper \cite{gmmpr2}.
 The lagrangian $L_1$ is defined in terms of the new hamiltonian,
\begin{equation}\label{e}
\begin{aligned}
H=&\int_{\Sigma} \sqrt{w}
d\sigma^{2}[\frac{1}{2}(\frac{P_{m}}{\sqrt{W}})^{2}
+\frac{1}{2}(\frac{\Pi^{r}}{\sqrt{W}})^{2}+\frac{1}{4}\{X^{m},X^{n}\}^{2}+\frac{1}{2}(\mathcal{D}_{r}X^{m})^{2}\\
& \nonumber
+\frac{1}{4}(\mathcal{F}_{rs})^{2}
] +
\Lambda(\{\frac{P_{m}}{\sqrt{W}},X^{m}\}-\mathcal{D}_{r}(\frac{\Pi^{r}}{\sqrt{W}})
]\\
\nonumber &+\int_{\Sigma} \sqrt{W}[-\overline{\Psi}\Gamma_{-}\Gamma_{r}\mathcal{D}_{r}\Psi+ \overline \Gamma_{-}\Gamma_{m}\{X^{m},\Psi\}]+\Lambda \{\overline{\Psi}\Gamma_{-}, \Psi\}.
\end{aligned}
\end{equation}

 {\bf Summarizing:} we departed with a theory whose degrees of freedom were $X^m$ scalars, $\Psi$ fermions, $dX^r$ closed forms, a global symmetry $Sp(2,Z)$, $N=2$ target supersymmetry, and a gauge symmetry: the $Diff_0(T^2)$, we have ended with a theory whose degrees of freedom are: $X^m$ scalars, $\psi_s$ fermions with $N=1$ target-space susy and r-gauge fields $A_r$ with gauge symmetry $Symp(\Sigma_g)$ and a global discrete symmetry $Z_2\times Z_2$. The $Z_2\times Z_2$ group is the Klein group isomorphic to the Dihedral symmetry group $Dih(2)$. The nontrivial fiber bundle is characterized by a integer $n\ne 0$ labeling the sum of the positive roots $\alpha_i$ associated to the complection of marked Riemann surface $\overline{\mathcal{M}}_{g,\alpha_i}$ for $g\ge 1$, $\alpha_i\ge 2 $. See \cite{monopole}\footnote{We thank R. Gopakumar for useful discussion to this respect.}. For genus $g=1$ the Riemann surface corresponds to the closure of the marked torus with at least two punctures, this is not the flat torus since the manifold has a nontrivial  negative curvature and corresponds to the $\frac{\mathcal{H}}{\Gamma_k}$  with $\mathcal{H}$ hyperbolic 2D space quotient in general by a family of of discrete symmetries $\Gamma_k$ associated to the particular gauging we are considering. More details will appear in \cite{tesisjoselen}. In this sense the supermembrane with central charge can be considered a gauging of the compactified supermembrane, in which the gauge field has been 'extracted' from the closed forms via a nontrivial fibration of topological  condition that has been partially gauged.

\section{On Further Applications}
In this section we  discuss qualitatively about the guidelines for further applications of this mechanism of gauging:

The most straightforward application, in the context of String Theory, would correspond to consider instead of a M2, a compactified D2-brane on Riemann surfaces in the presence of monopoles induced by the condition \cite{mrt}, by considering the complete Dirac-Born-Infeld action, or even for a stack of D2-branes. The general guidelines would correspond to those of the mechanism previously explained in section 3. The main difference relies in the absence of Poisson brackets, so the model should be conveniently adapted it. The non trivial flux quantization condition guarantees the existence of a new one-form connection in the compactified action. However to describe the full gauged action, one should analyze carefully how the global symmetries  are incorporated in the new gauged action. A nontrivial point is the construction of a proper covariant derivative for a deformed DBI lagrangian. A detailed analysis of this example is out of the scope of the present note.

 For the case of theories whose lagrangian is invariant under p-brane worldvolume diffeomorphisms, the generalization is rather straightforward. Let us consider a compactified p-brane $p\ge 2$ as a base manifold $B$, embedded on a compactified target space $F$. The usual compactification corresponds to the trivial fibration. The lagrangian $L_0$ corresponds to the gauge invariant functional of the sections. One can impose a topological condition to produce an obstruction for the triviality of the fibration. The condition required has to be such that it guarantees the existence of a 1-form on a principal bundle associated to the 2-cycle\footnote{For a topological interpretation in terms of embeddings, as an extra restriction, the 2-cycle has to be constructed via two 1-cycles.}. Once the gauge invariant functional (Lagrangian) has been constructed, the gauge invariant quantities exist in the complete fibration $E_1$. In the case of a line bundle this is just the curvature $F_2$. For higher p-forms the on has to follow the standard procedure to construct the appropriate invariants\footnote{$\approx Tr(F\wedge *F$)}. By considering base manifolds containing  a proper 2-cycle (constructed in terms of one-cycles),where to produce a nontrivial 2-cocycle in the Cech cohomology $H^2(B,\mathbb{Z})$ (discrete torsion) that guarantees the flux quantization condition to produce the nontrivial principal fiber bundle. One example is the central charge condition found in \cite{mrt} we showed in section 3, for the 2-cycles contained in the compactified manifold. This will guarantee the existence of a 1-form connection on the fiber bundle associated to the 2-cycle. This are particularly important when one is interested in incorporating nonabelian symmetry interactions in the deformed Lagrangian. This imposes the existence of a compatibility among the symplectic nondegenerate symplectic forms that restricts the gauge symmetry of the symplectomorphims group of the manifold $E$, by the monodromy group. A symplifying case could be to also consider the existence of 2-closed form inside a two cycle with the central charge condition, in such a way that a 1-form connection will also be defined via \bea
\int_{\Sigma_2} dX^r\wedge\ dX^s=\int_{\Sigma_2}F_2=n'\ne 0
\eea
and repeating the previous procedure.
All the subtleties with respect to the obstruction for the symplectic 2-form to be extended to the complete fibration as well as to see the compatibility between the different rank p-forms is outside the scope of this section but they should be considered carefully for particular constructions.\newline

 However it can be generalize more, by considering generalized version of this topological condition \cite{mrt} of irreducible wrapping for $p$ dimensions with $p>2$ \cite{gerbes}. Here we will concentrate the discussion just in the case of compactifications with a  generalized irreducible wrapping condition. This condition, -as also happens in the case of the central charge condition-, corresponds to a particular type of worldvolume flux condition.  So, these type of fluxes have a topological origin associated to the particular embedding condition of the p-brane in the compactified target space. The rest of the properties of the gauging mechanism should be carefully considered.

According to \cite{hoppe} the lagrangian of a  relativistic M-brane moving in D-dimensional space time may be
described, in a light-cone gauge,  by the VDiff$\Sigma$-invariant
sector of ([4])
\begin{equation}\label{abc}
H\;=\;\frac 1 2 \;\int_\Sigma\;\frac{d^M\varphi}{\rho (\varphi)}\;(\vec
p^{\;2} + g)
\end{equation}
where $g$ may be written in terms of Nambu-Poisson brackets,
\begin{equation}
g\;=\;\sum_{i_1<i_2<\cdots<i_M}\{ X_{i_1} ,\cdots, X_{i_M}\}
\{X^{i_1}, \cdots, X^{i_M}\},
\end{equation}
 $\{\cdots\}$ are the `Nambu-bracket' \footnote{The regularized version of these type of hamiltonians corresponds to the Fillipov algebras of $p$ degree. In the last times  3-algebras have received a considerable attention in terms of the AdS4/CFT3 realizations.}
 defined for scalar functions
$f_1,\cdots,f_M$ on $\Sigma$ as
\begin{equation}
\{f_1,\cdots,f_M\}\; :=\; \epsilon^{r_1\cdots r_M}\;\partial_{r_1}\;f_1
\cdots \partial_{r_M}\;f_M.
\end{equation}

When these M-branes are embedded in a compactified target space $M_{D-p}\times \mathcal{Y}_{p}$ appear maps subject to the winding condition
\bea
\begin{aligned}
&\oint_{C_s} dX^r=n_s^r\in \mathbb{Z}\quad C_s \quad \textrm{the homological basis of}\quad Im(\Sigma_p)\subset\mathcal{Y}\\
 &\oint_{M_{D-p}} dX^m=0
 \end{aligned}
 \eea  The $n_r^s\in Diff^+(\Sigma_p,\mathbb{Z})$. This is the same condition that appears in the case of the compactified supermembrane but particularized for the p-brane worldvolume $\Sigma_p$. The scalar functions $f_r$ are maps $(X^m,X^r)$ such that $X^{m,r}(\Sigma^i,\tau)$ depend of the $i$ spatial coordinates of the worldvolume base manifold. In two dimensions, the area preserving
diffeomorphisms are the same as the symplectomorphisms. In higher
dimensions,  the symplectomorphisms are a subgroup of the full
volume preserving diffeomorphisms. Now we impose what we will call a {\em generalized irreducible wrapping condition}\footnote{This condition was formerly analyzed in \cite{gerbes}.}

\bea
\int_{\Sigma_p} dX^1\wedge\dots\wedge dX^p=\int_{\Sigma_p}F_p=n\ne 0
\eea
So we have imposed a nonvanishing contribution of fluxes along the p-cycles contained in the compactified manifold assuming the base manifold $B$ to have no torsion\footnote{We thank to A. Restuccia for clarifying discussion with respect to these points.}. This defines a principal fiber bundle. There exists a gauge potential $A_{p-1}$ such that $F_p=dA_{p-1}$ is an $p-1$ form invariant under abelian symmetries,
\bea
A_{p-1}\to A_{p-1}+d\xi_{p-2}.
\eea
Derivatives of higher rank potentials can also be constructed although are more involved,  but examples have already been studied in the literature \cite{brahic}. So far the gauging procedure needs the existence of the monodromy representation for the mapping class group of arbitrarily p-dimensional surfaces, by defining it via choosing a representative.
The compactified manifold $\mathcal{Y}_k$ can have a dimension $k\ge p$ where $p$ denotes the worldvolume dimension. For $p$ even dimensional, the associated p-form will be nondegenerate on the fiber, and needs a careful study for its extension to the complete fibration $E$.  This condition, as happens for the case of the supermembrane, imposes a obstruction to the undeformed fibration (we assumed here a trivial one just associated to ordinary compactifications) which becomes non trivial, in the 'deformed' fibration. We conjecture that the resulting theory restricted by this topological condition is a gauged theory with respect to the associated to ordinary compactifications once the monodromy representative has been properly incorporated. The gauged theory is expected to contain global solutions possibly relevant to characterize interesting quantum properties as happens for the case of the supermembrane.
\newline

Take for example the M5-brane Hamiltonian (\ref{hp})
We start recalling the M5-brane Hamiltonian for the bosonic sector
in the light cone gauge that was obtained in
\cite{alexandra},
\begin{equation}\label{hp}
{\ch}_p=\frac{1}{2}\Pi^M\Pi_M+2g+l^{\mu \nu}l_{\mu\nu}+\Theta_{5i}\Omega^{5i}+\Theta_j\Omega^j
+\Lambda^{\alpha\beta}\Omega_{\alpha\beta},
\end{equation}
where
\begin{equation}\label{l}
l^{\mu\nu}=\frac{1}{2}(P^{\mu\nu}+\frac{\epsilon^{\mu\nu\gamma\delta\lambda}}{6}(\partial_{\rho}B_{\lambda\sigma})
+\p_{\sigma}B_{\rho\lambda} + \p_{\lambda}B_{\sigma\rho})
\end{equation}
$\Theta_{5i}$, $\Theta_j$, $\Lambda^{\alpha\beta}$ are the
Lagrange multipliers associated to the remaining constraints
(The two first constraints are the first class
constraints that generate the gauge symmetry associated to the
antisymmetric field and the third one is  the volume preserving
constraint. $P^{\mu\nu}$ and
$\Pi_M$ are the
conjugate momenta to $B_{\mu\nu}$ and ${X}^M$, respectively.
The elimination of second class constraints from the formulation in
\cite{Pasti2} and the third one is the responsible for producing a canonical Hamiltonian with
only first class constraints, was achieved at the price of loosing the
manifest 5 dimensional spatial covariance. In this way , the spatial
world volume splits into $M_{5}= M_{4}\times M_{1}$. The
supersymmetric version of this theory was given in  \cite{alexandra}.
 It may be expressed
directly in terms of the Nambu-Poisson bracket in five dimensions.
with $g$.
In \cite{m5} it was shown that $M_4$ admitted a symplectic
structure denoted as $\omega^0$. The scalar density $\sqrt{W}$, as for the case of the supermembrane with central charge
was also identified with the one arising from the symplectic structure over $M_4$.
By performing several partial gauge fixing on $B_{\mu\nu}¥$, following
\cite{alexandra} as a consequence of the Darboux's theorem , to express
$\omega_{kl}$ in terms of the  two-form $\omega^{0}$.

After fixing $\omega$ to $\omega^0$ we may resolve the
volume-preserving constraint for $\phi_{a}$ $a=1,2,3$. We are then
left still with one constraint,
\begin{equation}
\epsilon^{ijkl} \omega^0_{kl}\partial_i ( \frac{\Pi_M\partial_jX^M}{\sqrt{W}}) =0.
\end{equation}
Moreover, in the case of a non degenerate $\omega$, the second
Nambu-Poisson bracket may be re-expressed, on any open set of a Darboux atlas,
in terms of a Poisson bracket constructed with the symplectic two-form
$\omega^0_{ij}$ by fixing the volume preserving
diffeomorphisms:
In this case, the Hamiltonian is still invariant under the
symplectomorphisms which preserves $\omega^0$. We are then left with a
formulation in terms of $X^M$ and its conjugate momenta $\Pi_M$,
invariant under symplectomorphisms. The antisymmetric field
$B_{\mu\nu}$ and its conjugate momenta $P^{\mu\nu}$ have been
reduced to $\omega^0$, there is no local dynamics related to them. All
the dynamics may be expressed in terms of $(X^M,\Pi_M)$. We may
then perform the explicit $4+1$ decomposition  on the spatial
sector of the world-volume.
The determinant of the induced metric may be re-expressed in a
straightforward manner as a bracket
\begin{eqnarray}
g&=&
\frac{1}{5!}\{X^M,X^N,X^P,X^Q,X^R\}^2= \frac{64}{5!}(
\partial_5X^{[M} \{X^{[N},X^P\}\{X^{Q]},X^{R]}\})^2.
\end{eqnarray}
where the brackets on indices denote cyclic permutation.
All the interacting terms of the Hamiltonian (\ref{hp})
can then be expressed in terms of the Poisson bracket.
\newline
Now one can consider to embed it in a compactified target space. Obviously the embedding will modify the associated lagrangian density by incorporating the associated compactified sector. Once the flux associated to the topological condition is imposed
\bea
\int_{\Sigma_p}F_p=n\ne 0
\eea
naturally $\omega$ will be non-degenerate.
 the associated $\omega$ 2-form becomes symplectic on the nontrivial fibration. The existence of a nontrivial principal fiber bundle is not a sufficient condition to guarantee the existence of the gauged action as we have emphasized several times along the text.
The rest of the symmetries should be properly taken into account to incorporate to produce the gauged action according to the rules exposed previously.
We summarize then different steps of the mechanism:
 Take a model that contains a global symmetry group $H$, several closed one forms $\phi_r$ over a compact base manifold. The compact base manifold preserves the general coordinate transformation symmetry. In the case in which the base is a manifold with spatial coordinates, the symmetry preserved is a gauge symmetry of the infinite group of diffeomorphims preserving the volume. In 2 dimensions case, these are the group of area preserving diffeomorphims and they have associated a natural symplectic form on it. The base also posseses a particular isometry group $I$.
 \begin{itemize}
 \item{} Perform a Hodge decomposition separating the harmonic piece from the exact contribution.
 \item{Choose a metric defined on the base manifold in terms of the harmonic one-forms of the \textit{Fiber bundle} instead of $2g$ harmonic forms associated to the natural homology cycles of the base. (This represents a extra compatibility condition between the symplectic form of the fiber and the one on the base)}
 \item{}Incorporate the harmonic basis in the definition of a rotated covariant global derivative of a Weyl bundle. This fixes the harmonic function symmetry group $G$ by giving a scale to the Weyl covariant derivative of the bundle. These functions which are constants over the base, fix partially the global symmetry to a subgroup $\Gamma in H$.\\
\item{} Associate the exact one-form to a connection defined on the fiberbundle. Define a diffeomorphic connection $A$ preserving the fiber $F$. Define also a covariant derivative invariant under the gauge symmetry.\\
 \item{} Some terms of the action are canceled due to the new global properties associated to the nontrivial fibration. This properties define a global covariant derivative which is invariant under the residual global symmetry group $\Gamma$ in distinction with the usual one.\\

\item{} The projection of the connection on the base manifold defines a form preserving the diffeomorphisms of the base manifold. This projected one-form is not the canonical one of the undeformed base manifold $B$. Moreover since the isometry group of the base manifold has also changed due to the gauge fixing condition, we can interpret it as iff the compatibility of the fiber bundle implies a deformation of the original compact base $B_0$ to a new one $B_1$ but preserving its homotopy.

 \end{itemize}
 Finally we have ended with a theory that has a remanent discrete symmetry $\Gamma$, a new gauge symmetry which is a new kind of symplectomorphism (defined via $\Gamma$-invariant symplectic form in the bundle) in distinction with the original one that had a global symmetry group $H$ and a gauge symmetry group the canonical symplectomorphims of the original base. The number of degrees of freedom is kept constant, only closed 1-forms are "`converted" into a one-form connections, but fixing the harmonic degrees of freedom and the price for it, has been a change in the topology producing a residual invariance under global symmetries and but also in the geometry, a new gauge symmetry due to a change in the symplectomorphisms gauge symmetry of the associated fiber bundle to the principal $U(1)$ symmetry. If, for example, a trivial fibration is imposed, this does not allow to include the non-constant harmonic forms in the definition of the covariant derivative, and then one can not consistently extract the new gauge degrees of freedom. Moreover, when one tries to quantize the theory that has closed forms, one has to say how to deal with the well-known problem of quantizing harmonic forms.

\section{Conclusion}
The main conclusion is the following one: We have provided a new mechanism to gauge a theory corresponding to a nontrivial fibration over a compact base whose gauge field is extracted via the Hodge decomposition. The harmonic piece is properly handled in such a way that appears  discrete global symmetries associated to this sector. This method of gauging preserves the number of degrees of freedom of the undeformed theory. As an important example, we show that the supermembrane with central charges corresponds to a supermembrane compactified on a torus whose gauging is given by a SL(2,Z) fiber bundle with nontrivial monodromy.
We conjecture that the formulation of the supermembrane in terms of
sections of the symplectic torus bundle with a monodromy  is the natural way to understand the M-theory origin of the gauging procedures in supergravity theories \cite{gmmpr} and its low energy limit corresponds to the 9D $SL(2,\mathcal{R})$ gauged supergravities.. Coming back to our metaphor of the modeling clay, the gauged supergravities become gauged via the Noether type of gauging. However we consider that at High Energies the natural framework does correspond to the sculpting gauging method for the M2´s we have explained. It leads to the supermembrane with central charge theories.  As happens in our metaphor, when clay is cooked at high temperatures the only way to obtain a gauged structure, is with a chisel sculpting it.
Monodromies  associated to $\Gamma \subseteq H(\mathbb{Z})$ are fundamental emergent ingredients of this mechanism and they are naturally contained. At low energies the gauging group of the supergravities theories corresponds to the $G(\mathbb{R})$.
Let us make the following diagram for the diagram we propose with the two types of gaugings: 'sculpting' and Noether and their relations:
\begin{equation}
\begin{diagram}
\node{\textrm{Compactified M2($n=0$)} } \arrow[2]{e,t}{\textit{'Sculpting'}}
     \arrow{s,l}{\textrm{Low Energies}}
        \node[2]{\textrm{M2 with central charges ($n\ne 0$)}} \arrow{s,r}{\textrm{Low Energies}}\\
\node{\textrm{Maximal Supergrav. (d)}}\arrow[2]{e,t}{\textit{Noether}}
   \node[2]{\textrm{Gauged Supergravities (d)}}
		\label{eq:diag2}
\end{diagram}
\end{equation}

The precise relations with all details, are going to appear in \cite{gmpr2} and are part of \cite{tesisjoselen}.
\section{Acknowledgements}
 We thank to L. Alvarez-Gaum\'e, A. Lozada, J.F. Nu\~nez and  J.M. Pe\~na for interesting comments. To A Vi\~na we are particularly indebted, for all his kind explanations and clarifying discussions along the preparation of this draft, and to A. Restuccia  for a careful reading on the manuscript and for his fundamental help and discussions all along these years. We would also like thank to the Benasque Center for Science Pedro Pascual for kind hospitality during the last stages of completion of this paper and for the stimulating environment during the String Theory Workshop 2011. The work of MPGM is funded by the Spanish Ministerio de
Ciencia e Innovaci\'on (FPA2006-09199) and the Consolider-Ingenio
2010 Programme CPAN (CSD2007-00042).


\begin{thebibliography}{99}

\bibitem{cremmer-julia} E. Cremmer, B. Julia, Joel Scherk, {\em Supergravity Theory in Eleven-Dimensions.}
 Phys.Lett.{\bf B76}:409-412,1978.
 %
\bibitem{bl} Bailin and Love, {\em Supersymmetric Gauge Field Theory and String Theory} Graduate Student Series in Physics, IOP.
%
\bibitem{dwn}B. de Wit, H. Nicolai, {\em N=8 Supergravity with Local SO(8) x SU(8) Invariance.}
 Phys.Lett.{\bf B108}:285,1982.
%
\bibitem{hull} C.M. Hull, {\em Noncompact Gaugings Of N=8 Supergravity.}
Phys.Lett.{\bf B142}:39,1984.
%
\bibitem{samtleben} H. Samtleben, {\em Lectures on Gauged Supergravity and Flux Compactifications.} Class.Quant.Grav.{\bf 25}:214002,2008.
{\tt arXiv:0808.4076 [hep-th]}.
%
\bibitem{mrt} I. Martin, A. Restuccia, R. S. Torrealba {\em On the stability of compactified D = 11 supermembranes.}
 Nucl.Phys. {\bf B521}:117-128,1998.{\it hep-th/9706090}.
%

\bibitem{mor} I. Martin, J. Ovalle, A. Restuccia, {\em D-branes, symplectomorphisms and noncommutative gauge theories}, Nucl. Phys. Proc. Suppl. {\bf 102}(2001) 169-175. {\em Compactified D = 11 supermembranes and symplectic noncommutative gauge theories}, Phys. Rev. {\bf D64} (2001) 046001, {\tt hep-th/0101236}.
%
\bibitem{bellorin} J. Bellorin, A. Restuccia, {\em D=11 Supermembrane wrapped on calibrated submanifolds}, Nucl. Phys. {\bf B737} 190-208, 2006,{\tt hep-th/0510259}.
%
\bibitem{joselen} M. P. Garcia del Moral, J. M. Pena, A. Restuccia {\em N=1 4D Supermembrane from 11D}
   JHEP0807:{\bf 039},2008, {\tt arXiv:0709.4632}.
%
\bibitem{gmmr} M.P. Garcia del Moral, I. Martin, A. Restuccia, {\em Nonperturbative SL(2,Z) (p,q)-strings manifestly realized on the quantum M2.} {\tt arXiv:0802.0573 [hep-th]}

\bibitem{g2} A. Belhaj, M.P. Garcia del Moral, A. Restuccia, A. Segui, J.P. Veiro,{\em The Supermembrane with Central Charges on a G2 Manifold.} J.Phys. {\bf A42}:325201,2009. {\tt arXiv:0803.1827}.
%
\bibitem{gmmpr} M.P.Garcia del Moral, I. Martin, J.M. Pena, A. Restuccia, {\em SL(2,Z) symmetries, Supermembranes and Symplectic Torus},
 {\tt arXiv:1105.3181 [hep-th]}.
 %
 %
\bibitem{gmpr2} M.P. Garcia del Moral, J.M.Pena. In preparation.
%
%
\bibitem{dwhn} B. de Wit, J. Hoppe, H. Nicolai, {\em On the quantum mechanics of
supermembranes}. Nucl. Phys. {\bf B305}: 545,1988.
%
\bibitem{dwln} B. de Wit, M. Luscher, H. Nicolai, {\em The supermembrane is unstable}.
Nucl. Phys. {\bf B320}: 135, 1989.
%
\bibitem{dwmn} B. de Wit, U. Marquard, H. Nicolai,
{\em Area preserving diffeomorphisms and supermembrane lorentz
invariance.} Commun. Math. Phys. {\bf 128}:39-62, 1990.
%
\bibitem{dwpp} B. de Wit, K. Peeters, J. Plefka,
{\em Supermembranes with winding.} Phys. Lett.{\bf B409}: 117-123,
1997. {\it hep-th/9705225}
%
%
\bibitem{gmr} M.P. Garcia del Moral, A. Restuccia, {\em On the spectrum of a noncommutative formulation of the D=11 supermembrane with winding}, Phys. Rev. {\bf D66} (2002) 045023, {\tt hep-th/0103261}.
%
\bibitem{bgmr}
L. Boulton, M.P. Garcia del Moral, A. Restuccia, {\em Discreteness of the spectrum of the compactified D=11 supermembrane with non-trivial winding}, Nucl. Phys. {\bf B671} (2003) 343-358, {\tt hep-th/0211047}.
%
\bibitem{bgmr2}
L. Boulton, M.P. Garcia del Moral, A. Restuccia, {\em The Supermembrane with central charges: (2+1)-D NCSYM, confinement and phase transition}, {\tt hep-th/0609054}.
%
\bibitem{bgmr3} L.Boulton, M.P. Garcia del Moral, A. Restuccia,{\em Spectral properties in supersymmetric matrix models.}
 arXiv:1011.4791 [hep-th].

\bibitem{kravchenko} O. Kravchenko, {\em Deformation Quantization of Symplectic Fibrations}. Compositio Mathematica
Volume 123, Number 2, 131-165, DOI: 10.1023/A:1002452002677.
%
\bibitem{alvarez} O. Alvarez {\em Topological quantization and cohomology} Comm. Math.Phys.{\bf 100} 279-309, 1985.


\bibitem{ash} A. Ash, {\em Farrell cohomology of $GL(n,Z)$},Israel Journal of Mathematics, {\bf 67},3,1989.
%
\bibitem{gmx} H.H. Glover, G. Mislin, Y. Xia {\em On the Farrell cohomology of mapping class group}Inventiones Mathematicae
Volume 109, Number 1, 535-545, 1992.
%
\bibitem{feng-luo} Feng-Luo {\em Torsion elements in the mapping class group of a surface} arXiv: math/0004048.
%
\bibitem{khan} P. J. Khan, {\em Symplectic torus bundles and group extensions}. New York Journal of Mathematics. New York J. Math. {\bf 11}: 35–55, 2005.
    %
\bibitem{walzack} R. Walczak, {\em Existence of symplectic structures on torus bundles over surfaces}. {\tt
math/0310261}.
%
\bibitem{weyl} A. Weil,Variétés Kaehlériennes, Hermann (1957).
%
\bibitem{mr} I. Martin, A. Restuccia, {\em Symplectic connections, noncommutative Yang-Mills theory and supermembranes.}
 Nucl.Phys.{\bf B622}:240-256,2002.{\it hep-th/0108046}.
%
\bibitem{monopole} I. Martin, A. Restuccia, {\em  Magnetic monopoles over topologically nontrivial Riemann surfaces.}
Lett.Math.Phys.{\bf 39 }:379-391,1997. {\it hep-th/9603035}.
%
\bibitem{gmmpr2}M.P.Garcia del Moral, I. Martin, J.M. Pena, A. Restuccia. In preparation.
%
\bibitem{tesisjoselen} Joselen M.  Pena, Ph.D Thesis, Current.
%
\bibitem{gerbes} M.I. Caicedo, I. Martin, A. Restuccia,{\em Gerbes and duality},
 Annals Phys.{\bf 300}:32-53,2002, {\tt hep-th/0205002}.
%
\bibitem{hoppe} Jens Hoppe, {\em On M algebras, the quantization of Nambu mechanics, and volume preserving diffeomorphisms.}
Helv.Phys.Acta {\bf 70}:302-317,1997. {\it hep-th/9602020}.
%
\bibitem{brahic} O. Brahic {\em On the infinitesimal gauge symmetries of closed forms} arXiv:1010.2189.
%
\bibitem{alexandra} A. De Castro, (Caracas, IVIC) , A. Restuccia, {\em Superfive-brane Hamiltonian and the chiral degrees of freedom.}
 Phys.Rev.{\bf D66}:024037,2002. {\tt hep-th/0204052}.
%

\bibitem{Pasti2} I.Bandos, K. Lechner, A. Nurmagambetov, P. Pasti, D. Sorokin, and M. Tonin, {\em Covariant action for the superfive-brane of M theory.} Phys.Rev. Lett.,{\bf 78}:4332, 1997. {\tt hep-th/9701149}.
%
\bibitem{m5} A. De Castro, M.P. Garcia del Moral, I. Martin, A. Restuccia,
{\em M5-brane as a Nambu-Poisson geometry of a multiD1-brane theory.}
 Phys.Lett.{\em B584}:171-177,2004. {\it hep-th/0306094}.






\end{thebibliography}
\end{document}